\documentclass{elsart}
\usepackage{epsfig}
\usepackage{natbib}
\begin{document}
\runauthor{Cicero, Caesar and Vergil}
\begin{frontmatter}
\title{Predictions of Alpha Decay Half lives of Heavy and Superheavy Elements}

\author[SINP,VCU]{C. Samanta\thanksref{Y}},
\author[SINP]{P. Roy Chowdhury\thanksref{X}}

\thanks[Y]{E-mail:chhanda.samanta@saha.ac.in}
\thanks[X]{E-mail:partha.roychowdhury@saha.ac.in}

\address[SINP]{Saha Institute of Nuclear Physics, 1/AF Bidhan Nagar, Kolkata 700 064, India }
\address[VCU]{Physics Department, Virginia Commonwealth University, Richmond, VA 23284-2000, U.S.A.}
and
\author{D.N. Basu\thanksref{Z}}
\address{Variable  Energy  Cyclotron  Centre, 1/AF Bidhan Nagar, Kolkata 700 064, India }
\thanks[Z]{E-mail:dnb@veccal.ernet.in}

\begin{abstract}

    Theoretical estimates for the lifetimes of several isotopes of heavy elements with Z~=~102-120 are presented by calculating the quantum mechanical tunneling probability in a WKB framework and using microscopic nucleus-nucleus potential obtained by folding the densities of interacting nuclei with the DDM3Y effective nuclear interaction. The $\alpha$-decay half lives calculated in this formalism using the experimental $Q$-values are in good agreement over a wide range of experimental data.  Half lives are also calculated using $Q$-values extracted from two mass formulae. The Viola-Seaborg-Sobiczewski (VSS) estimates of $\alpha$-decay half lives with the same $Q$-values are presented for comparison. The half life calculations are found to be quite sensitive to the choice of $Q$-values. Comparison with the experimental data delineates the inadequacies of older mass predictions in the domain of heavy and superheavy elements as compared to the newer one by Muntian-Hofmann-Patyk-Sobiczewski, and highlights necessity of a more accurate mass formula which can predict  $Q$-values with even higher precision.\\

\end{abstract}

\end{frontmatter}
\section{Introduction}

      Currently nuclear physics is passing through an exciting period of time when one after another new superheavy elements (SHE) are being discovered. The nuclear shell model predicts that the next magic proton number beyond Z~=~82 would be Z~=~114. But some recent microscopic nuclear theories suggest a magic island or, the $island~of~stability$  around  Z~=~120, 124, or, 126 and N~=~184~\cite{st06}. In this island there might be shell gaps or, a region of evenly spaced levels with no pronounced shell gap. A tremendous progress in experiments and accelerator technologies has made it possible to reach near the shore of this magic island. The heavy elements with Z~=~107~-~112 have been successfully synthesized at GSI, Darmstadt~\cite{ho04,ho00,ac06}. Isotopes of these elements along with Z~=~114~-~116 and 118 have been synthesized at JINR-FLNR, Dubna ~\cite{oga06,oga04,og04} and an isotope of Z~=~113 has been identified at RIKEN, Japan~\cite{mo04,MO04}. Now, the upcoming radioactive ion beam facilities in Japan, Europe and U.S.A.  promise to produce more neutron-rich heavy elements and, may be the ultimate SHE with the longest life time.  

A superheavy nucleus predominantly undergoes sequential $\alpha$-decays followed by subsequent spontaneous fission. In experiment one usually measures the decay energies and decay times, while one of the major goals of theory is to be able to predict the $\alpha$-decay lifetimes of SHE. This work delineates the usefulness and limitations of some recent mass formulae in predicting the alpha decay half lives, and points out the necessity of a more accurate mass formula in the heavy and superheavy region.

Earlier we investigated the predictive power of the alpha decay half life calculations in a WKB framework with DDM3Y interaction ~\cite{prc06} in which we used the experimental alpha decay Q-values of even Z nuclei only, from Z =106 - 118. The half lives calculated in this formalism were found to be in better agreement with the experimental data compared to the VSS~\cite{VSS89} predictions calculated with the same experimental Q values (Table I of ref. ~\cite{prc06}). Theoretical Q-values obtained from ref.~\cite{ms} and ref.~\cite{Sm97} were presented for comparison with the experimental Q-values. Predictions of the generalized liquid drop model (GLDM)~\cite{roy02,roy04} were also presented  and it was pointed out that the disagreements of the results with the experimentally observed half lives  were primarily due to use of theoretical Q values that differ from the experimental ones.

In Table II of ref.~\cite{prc06}, the $log_{10} T$ values for altogether 76 even-even isotopes of Z = 104 - 120, computed with the theoretical Q-values from the macroscopic-microscopic (M-M) model~\cite{Sm97}, were presented in the same WKB framework and compared with the VSS predictions and Viola-Seaborg estimates used in ref.~\cite{Sm97}.  In general, calculations in the WKB framework with DDM3Y interactions and experimental Q-values were found to be in better agreement with the experimental half lives, than those computed using other formalisms. 

In the present work we have carried out the half life calculations for both even and odd Z nuclei from 
Z = 106 - 118 with slightly different experimental Q-values obtained from a recent paper~\cite {oga06}, published after our previous work~\cite{prc06}. Moreover, we present DDM3Y and VSS estimates of half lives of altogether 314 even-even, even-odd, odd-odd and odd-even  isotopes of Z= 102 - 120 with theoretical Q-values extracted from ref.~\cite{ms} and a new mass formula by Muntian et al.~\cite{Mu01, MU03, Mu03}. 

Nuclear masses that have been obtained accurately within a statistical Thomas-Fermi model with a Yukawa plus exponential type phenomenological effective interaction~\cite{ms,Mo95} show appreciable deviations while predicting for the SHE region. Microscopic Hartree-Fock self-consistent calculations using mean fields and Skyrme or Gogny forces and pairing correlations~\cite{Sa02,Ri05} as well as relativistic mean field theories~\cite{Be01,ga05} have also been developed to describe these nuclear masses. Nuclear mass systematics using neural networks have also been undertaken recently~\cite{At04}. We find that a  macroscopic-microscopic model of  mass predictions by Muntian et al.~\cite{Mu01,MU03,Mu03}  to be slightly superior. However, the present investigation shows that in order to be able to provide accurate and fully theoretical estimates of decay lifetimes, a lot more is desired for the mass predictions in the domain of SHE.

Theoretical estimates for the lifetimes are provided by calculating the quantum mechanical tunneling probability in a WKB framework and using microscopic nucleus-nucleus potential. The microscopic nucleus-nucleus potentials are obtained by folding the densities of interacting nuclei with the DDM3Y effective nuclear interaction~\cite{Be77,sa79}. The $\alpha$-decay half lives calculated in this formalism using the experimental $Q$-values (extracted from the measured $\alpha$-particle kinetic energies) are in good agreement over a wide range of experimental data~\cite{oga06,oga04,og04,mo04}. The Viola-Seaborg estimates of $\alpha$-decay half lives with Sobiczewski constants  (VSS) and using the same $Q$-values are also presented for comparison. Such a comparison highlights the accuracy of the half life predictions for the present calculations with DDM3Y interaction and the calculations being sensitive to the $Q$-values, it allows to discriminate among various available mass predictions in the domain of heavy and superheavy elements. Although the macroscopic-microscopic mass formula of Muntian et al.~\cite{Mu01,MU03,Mu03} provides best estimates for the decay $Q$-values in the domain of SHE, but it still lacks the level of accuracy that the present calculations deserve for accurate half life predictions.      

      In section 2, a description of the formalism  is given. In section 3, the essential ingradients of the macroscopic-microscopic mass formula of Muntian et al.~\cite{Mu01,MU03,Mu03} are described.  In Section 4, the Viola-Seaborg-Sobiczewski approach for $\alpha$-decay half lives is briefly outlined. In section 5, the calculations and results and in section 6, the summary and conclusions are presented.

\vspace{-0.52cm}

\section{Formalism}

\vspace{-0.52cm}

    The  $\alpha$ decay half lives are calculated in the frame work of quantum mechanical tunneling of an alpha particle from a parent nucleus~\cite{prc06}. The required nuclear interaction potentials are calculated by double folding the density distribution functions of the $\alpha$ particle and the daughter nucleus with density dependent M3Y effective interaction. The microscopic $\alpha$-nucleus potential thus obtained, along with the Coulomb interaction potential and the minimum centrifugal barrier required for the spin-parity conservation, form the potential barrier. The half lives of  $\alpha$ disintegration processes are calculated  using the WKB approximation for barrier penetrability. Spherical charge distributions have been used for calculating the Coulomb interaction potentials. The $Q$-values of $\alpha$-decay are obtained from both the experimental data and theoretical predictions.

\subsection{The $Q$-values of $\alpha$-decay}

      The experimental decay $Q$ values ($Q_{ex}$) have been obtained from the measured $\alpha$-particle kinetic energies $E_{\alpha}$ using the following expression 

\begin{equation}
 Q_{ex} = (\frac{A_p}{A_p-4})E_{\alpha} + (65.3 Z_p^{7/5} - 80.0 Z_p^{2/5}) \times 10^{-6} ~\rm MeV
\label{seqn1}
\end{equation}
\noindent
where the first term is the standard recoil correction and the second term is an electron shielding correction in a systematic manner as suggested by Perlman and Rasmussen \cite{Pe57}. 

      The theoretical decay $Q$ values $Q_{th}$ have been obtained from theoretical estimates for the atomic mass excesses~\cite{ms,Mu01,MU03,Mu03} using the following relationship 

\begin{equation}
 Q_{th} = M - ( M_\alpha + M_d) = \Delta M - (\Delta M_\alpha + \Delta M_d)
\label{seqn2}
\end{equation}
\noindent
which if positive allows the decay, where $M$, $M_\alpha$, $M_d$ and $\Delta M$, $\Delta M_\alpha$, $\Delta M_d$ are the atomic masses and the atomic mass excesses of the parent nucleus, the emitted $\alpha$-particle and the residual daughter nucleus, respectively, all expressed in the units of energy.

\subsection{The microscopic nuclear potentials for the $\alpha$-nucleus interaction}

The nuclear interaction potential $V_N(R)$ between the daughter nucleus and the emitted particle is obtained in a double folding model as~\cite{sa79},

\begin{equation}
 V_N(R) = \int \int \rho_1(\vec{r_1}) \rho_2(\vec{r_2}) v[|\vec{r_2} - \vec{r_1} + \vec{R}|] d^3r_1 d^3r_2 
\label{seqn3}
\end{equation}
\noindent
where $\rho_1$ and $\rho_2$ are the density distribution functions for the two composite nuclear fragments and $v[|\vec{r_2} - \vec{r_1} + \vec{R}|]$ is the effective NN interaction. The density distribution function in case of $\alpha$ particle has the Gaussian form

\begin{equation}
 \rho(r) = 0.4229~{\rm exp}( - 0.7024 r^2)
\label{seqn4}
\end{equation}                                                                                                                                          
\noindent     
whose volume integral is equal to $A_\alpha ( = 4 )$, the mass number of $\alpha$-particle. The matter density distribution for the daughter nucleus can be described by the spherically symmetric Fermi function 

\begin{equation}
 \rho(r) = \rho_0 / [ 1 + {\rm exp}( (r-c) / a ) ]
\label{seqn5}
\end{equation}                                                                                                                                       
\noindent     
where the equivalent sharp radius $r_\rho$, the half density radius $c$ and the diffuseness for the leptodermous Fermi density distributions are given by
                        
\begin{equation}
c = r_\rho ( 1 - \pi^2 a^2 / 3 r_\rho^2 ),~~r_\rho = 1.13 A_d^{1/3},~~a = 0.54~fm
\label{seqn6}
\end{equation}
\noindent
and the value of the central density $\rho_0$ is fixed by equating the volume integral of the density distribution function to the mass number $A_d$ of the residual daughter nucleus. 

      The distance $s$ between any two nucleons, one belonging to the residual daughter nucleus and other belonging to the emitted $\alpha$, is given by $s = |\vec{r_2} - \vec{r_1} + \vec{R}|$ while the interaction potential between these two nucleons $v(s)$ appearing in eqn.(3) is given by the factorised DDM3Y effective interaction. The general expression for the DDM3Y realistic effective NN interaction used to obtain the double-folded nucleus-nucleus interaction potential is given by,  
\begin{equation}
  v(s,\rho_1,\rho_2,\epsilon) = t^{M3Y}(s,\epsilon)g(\rho_1,\rho_2)
\label{seqn7}
\end{equation}   
\noindent
where the isoscalar $t_{00}^{M3Y}$ and the isovector $t_{01}^{M3Y}$ components of M3Y interaction potentials \cite{sa79} supplemented by zero range potentials are given by the following equations: 

\begin{equation}
 t_{00}^{M3Y}(s, \epsilon) = 7999\frac{\exp( - 4s)}{4s} - 2134\frac{\exp( - 2.5s)}{2.5s} - 276 (1 - \alpha\epsilon)\delta(s)
\label{seqn8}
\end{equation} 
\noindent 

\begin{equation}
  t_{01}^{M3Y}(s, \epsilon) =  -4886\frac{\exp( - 4s)}{4s} + 1176\frac{\exp( - 2.5s)}{2.5s} + 228 (1 - \alpha\epsilon)\delta(s).
\label{seqn9}
\end{equation}   
\noindent
where $\epsilon$ is the energy per nucleon. 
The isovector term does not contribute if anyone (or, both) of the daughter and emitted nuclei involved in the decay process has  N=Z, N and Z being the neutron number and proton number respectively. Therefore in $\alpha$-decay calculations only the isoscalar term contributes. The density dependence term $g(\rho_1, \rho_2)$ can be factorized~\cite{prc06} into a target term times a projectile term as,

\begin{equation}
 g(\rho_1, \rho_2) = C (1 - \beta \rho_1^{2/3}) (1 - \beta \rho_2^{2/3})
\label{seqn10}
\end{equation}   
where C, the overall normalisation constant, is kept equal to unity and the parameter $\beta$ can be related to the mean-free path in the nuclear medium with value equal to 1.6 $fm^2$~\cite{prc06}. The $\rho_1$ and $\rho_2$ are the density distributions of the $\alpha$-particle and the daughter nucleus respectively.

\subsection{The $\alpha$-decay half lives of superheavy nuclei}

      The half life of a parent nucleus decaying via $\alpha$ emission is calculated using the WKB barrier penetration probability~\cite{prc06}.  The decay half life $T_{1/2}$ of the parent nucleus $(A, Z)$  into a $\alpha$ and a daughter $(A_d, Z_d)$  is given by,

\begin{equation}
 T_{1/2} = [(h \ln2) / (2 E_v)] [1 + \exp(K)]
\label{seqn11}
\end{equation}
\noindent
where $E_v$ is the zero point vibration energy. The zero point vibration energies used in the present calculations are $E_v$ = 0.1045$Q$ for even-even, 0.0962$Q$ for odd Z-even N, 0.0907$Q$ for even Z-odd N, 0.0767$Q$ for odd-odd parent nuclei and are the same as that described in ref. \cite{Po86} immediately after eqn.(4) which were obtained from a fit to a selected set of experimental data on $\alpha$ emitters and includes the shell and the pairing effects. The action integral $K$ within the WKB approximation is given by

\begin{equation}
 K = (2/\hbar) \int_{R_a}^{R_b} {[2\mu (E(R) - E_v - Q)]}^{1/2} dR
\label{seqn12}
\end{equation}
\noindent
where the total interaction energy $E(R)$ between the $\alpha$ and the residual daughter nucleus is equal to the sum of the nuclear interaction energy, Coulomb interaction energy and the centrifugal barrier. Thus

\begin{equation}
 E(R) = V_N(R) + V_C(R) + \hbar^2 c^2 l(l+1) / (2\mu R^2)
\label{seqn13}
\end{equation}   
\noindent
where the reduced mass $\mu = M_{\alpha} M_d/ (M_{\alpha} + M_d)$ and $V_C(R)$ is the Coulomb potential between the $\alpha$ and the residual daughter nucleus. $R_a$ and $R_b$ are the second and third turning points of the WKB action integral determined from the equations 

\begin{equation}
 E(R_a)  = Q + E_v =  E(R_b)
\label{seqn14}
\end{equation}
\noindent
whose solutions provide three turning points. The $\alpha$ particle oscillates between the first and the second turning points and tunnels through the barrier at $R_a$ and $R_b$. The zero point vibration energy $E_v$ appearing in the denominator of eqn.(11) is proportional to the released energy $Q$. Also, through the action integral [eqn.(12)],  $Q$ goes to the exponential function in eqn.(11). Therefore, the lifetime calculations become very sensitive to the released energies $Q$ involved in the decay processes.

\section{Macroscopic-microscopic mass formula in heavy mass region}

The ground state mass of a nucleus (A, Z) was calculated by Muntian et al.~\cite{Mu01,MU03,Mu03} within a macroscopic-microscopic approach.  The Yukawa-plus-exponential model \cite{kra79} was used for the macroscopic part and the Strutinski shell correction for the microscopic part. The macroscopic part \cite{Mu01} is given by 

\begin{eqnarray}
M_{macr}(Z, N, \beta^0_\lambda)=&&M_HZ+M_nZ-a_v(1-\kappa_vI^2)A       \nonumber\\              &&+a_s(1-\kappa_sI^2)A^{2/3}B_1(\beta^0_\lambda) +a_0A^0                     \nonumber\\ 
&&+c_1Z^2A^{-1/3}B_3(\beta^0_\lambda)-c_4Z^{4/3}A^{-1/3}                     \nonumber\\
&&+f(k_Fr_p)Z^2A^{-1}-c_a(N-Z)-a_{el}Z^{2.39}                                          
\label{seqn15}
\end{eqnarray}
\noindent
where $M_H$ is the mass of the hydrogen atom, $M_n$ is mass of neutron, I=(N-Z)/A is the relative neutron excess and A=Z+N is the mass number of a nucleus. The functions $B_1(\beta_\lambda)$ and $B_3(\beta_\lambda)$ describe the dependence of the surface and Coulomb energies, respectively, on the deformation $\beta_\lambda$, and $\beta^0_\lambda$ is the value of the deformation at equilibrium. The coefficient $c_1$ and $c_4$ and the function $f(k_Fr_p)$ have the same form and values as in \cite{mo81,Mo95}, with the proton root-mean square radius $r_p$=0.80~fm and the nuclear radius constant $r_0$=1.16~fm. The electron binding constant is $a_{el}=1.433\times10^{-5}$MeV \cite{mo81,Mo95}. The values of parameters $a_s$ and $\kappa_s$ of surface term are 21.13 MeV and 2.30 respectively. Adjusted values of the parameters $a_v$, $\kappa_v$, $a_0$, $c_a$ to experimental masses of heaviest nuclei ($Z >$ 83) are 16.0643, 1.9261, 17.926 and 0 respectively.

                In the microscopic part, adjustment of pairing force strength is used with isotopic-dependent form of the monopole type pairing strength $AG_l=g_{0l}+g_{1l}I$ which results $g_{0n}=17.67$ MeV, $g_{1n}=-13.11$ MeV, for $l=n$~(neutrons)
and $g_{0p}=13.40$ MeV, $g_{1p}=44.89$ MeV, for $l=p$~(protons). 
 
\vspace{-0.2cm}

\section{The Viola-Seaborg-Sobiczewski approach for $\alpha$-decay half lives}

\vspace{-0.2cm}

      The $\alpha$ decay half lives estimated by Viola-Seaborg semi-empirical relationship with constants determined by Sobiczewski, Patyk and Cwiok~\cite{VSS89} is given by

\begin{equation}
 log_{10} T_{1/2}  = [a Z + b] [Q/MeV]^{-1/2} + cZ + d + h_{log}
\label{seqn16}
\end{equation}
\noindent
where the half-life $T_{1/2}$ is in seconds, the $Q$-value is in MeV,  $Z$ is the atomic number of the parent nucleus. Instead of using original set of constants by Viola and Seaborg, more recent values 
\begin{equation}
 a=+1.66175,~~~~b=-8.5166,~~~~c=-0.20228,~~~~d=-33.9069
\label{seqn17}
\end{equation}
\noindent that were determined in an adjustment taking account of new data for new even-even nuclei~\cite{VSS89} are used. The quantity $h_{log}$ in eqn.(16) accounts for the hindrances associated with odd proton and odd neutron numbers given by Viola and Seaborg~\cite{Vi}, namely

\begin{eqnarray}
 h_{log} = &&~~0~~~~~~for~Z~even-N~even\nonumber\\
            = && 0.772~~for~Z~odd-N~even\nonumber\\
            = && 1.066~~for~Z~even-N~odd\nonumber\\
            = && 1.114~~for~Z~odd-N~odd
\label{seqn18}            
\end{eqnarray}   
\noindent       
The uncertainties in the calculated half lives due to this semi-empirical approach are far smaller than the uncertainties due to errors in the calculated energy release. 

\section{Calculations and Results}

      Calculations have been performed assuming spherical charge distribution for the residual daughter nucleus and the emitted nucleus as a point particle. The Coulomb interaction potential $V_C(R)$ between them is given by

\begin{eqnarray}
 V_C(R) =&&(\frac{Z_eZ_de^2}{2R_c}).[3-(\frac{R}{R_c})^2]~for~R\leq R_c, \nonumber\\
            = &&\frac{Z_e Z_d e^2}{R}~~otherwise
\label{seqn19}            
\end{eqnarray}   
\noindent
where $Z_e$ and $Z_d$ are the atomic numbers of the emitted-cluster and the daughter nucleus respectively. The touching radial separation $R_c$ between the emitted-cluster and the daughter nucleus is given by $R_c = c_e+c_d$ where $c_e$ and $c_d$ have been obtained using eqn.(6). 

      Comparison between experimental and calculated $\alpha$-decay half-lives for zero angular momenta transfers, using spherical charge distributions for Coulomb interaction and the DDM3Y effective interaction is provided in Table 1. The lower and upper limits of the theoretical half lives corresponding to the upper and lower limits of the experimental $Q_{ex}$ values are also provided. The quantitative agreement with experimental data is reasonable. The results which are underestimated are possibly because the centrifugal barrier required for the spin-parity conservation could not be taken into account due to non availability of the spin-parities of the decay chain nuclei. The term $\hbar^2 c^2 l(l+1) / (2\mu R^2)$ in eqn.(13) represents the additional centrifugal contribution to the barrier that acts to reduce the tunneling probability if the angular momentum carried by the $\alpha$-particle is non-zero. Hindrance factor which is defined as the ratio of the experimental $T_{1/2}$ to the theoretical $T_{1/2}$ is therefore larger than unity since the decay involving a change in angular momentum can be strongly hindered by the centrifugal barrier. 
\noindent


%

       To study the predictive power of the mass formula, $Q$-values are also calculated using the mass formula of Myers and Swiatecki [$Q^{MS}_{th}$] and Muntian et al.  [$Q^{M}_{th}$]. Experimental half lives are given only in Table 1. Comparison with the theoretical half life values indicates that the $Q$-value predictions of Muntian et al. for the SHE domain are in better agreement with the experimental data than the values obtained from the Myers-Swiatecki mass table.  For example, the theoretical half life of $^{289}114$ obtained using the $Q^{MS}_{th}$  is $\sim 700$ times the experimental one. Where as, calculations in the same framework but with [$Q^{M}_{th}$] as well as experimental $Q_{ex}$-values, agree well over a wide range of experimental data. The theoretical  VSS estimates for $T_{1/2}$ largly overestimates in many cases showing inconsistencies while the present estimate is inconsistent only for few cases where it overestimates but still provides much better estimate than that estimated by the VSS systematics.

The Table 1 shows that the $T_{1/2}$-value decreases as $Q$ increases. But, this predominant trend is not observed in the experimental data of the  $^{274}111$ and $^{270}109$~\cite{mo04}, origin of which can be ascertained once spin-parity assignments of such odd-odd nuclei and their decay chain  nuclei are known.

Figures 1-5 show theoretical estimates for the $\alpha$-decay half lives of 314 heavy and superheavy elements. The theoretical $Q$-values have been computed using eqn.(2) and the mass formulae of Myers-Swiatecki [MS]~\cite{ms} and  Muntian et al. [M]~\cite{Mu01,MU03,Mu03}. The figures show that the alpha decay half lives of neutron rich Z~=~102, N~=~162 ($T_{1/2}[Q^M_{th}]~=~3.06~\times 10^{+09}$ s) is more than the elements of higher Z considered in the present study. For Z~=~115, 116, 117, 118, 119 and 120, the variation of $T_{1/2}$ with the neutron number becomes apparently flat at larger N values and then drops after N~=~184 indicating a possible signature of neutron number magicity at N~=~184. For lower Z values, Q values beyond N~=184 are not available from ref.~\cite{Mu01, MU03,Mu03}. 

 Interestingly, although N~=~184 is predicted to be a magic number~\cite{st06}, for Z~=~114~-~120, the N~=~183 isotopes are found to have longer half lives than the N~=~184 isotopes. In fact some other isotopes in this domain are predicted to have even longer half lives. For example in the range of $^{282}114 - ^{298}114$, the $^{297}114$  isotope has the highest $T_{1/2} $ value and, in the range of $^{290}118 - ^{306}118$, the $^{293}118$  isotope has the highest $T_{1/2}$ value ($T_{1/2}[Q^M_{th}]$ of $^{302}118$ and $^{293}118$  are $ 2.58~\times 10^{-04}$ and $8.93~\times 10^{-04}$ seconds, respectively). 

The heavy element Z~=~120, N~=~183 isotope does not show any extra stability compared to the N~=~183 isotopes of Z~=~114~-~119. For example, according to the macroscopic-microscopic mass formula of Muntian et al., the half life sequences for N=183 isotopes are: $^{297}114 ~(T_{1/2}~=~1.47~\times~10^3~s$),  $^{298}115 ~(T_{1/2}~=~6.78~\times~10^2~s$),  $^{299}116 ~(T_{1/2}~=~1.84~\times~10^{-1}~s$),  $^{300}117 ~(T_{1/2}~= 2.46~\times 10^{-2}~s$),  $^{301}118 ~(T_{1/2} = 6.45~\times~10^{-4}~s$),  $^{302}119~(T_{1/2}~=~1.63~\times~10^{-4}~s$),  $^{303}120 ~(T_{1/2}~=~1.27~\times~10^{-5}~s$). 

\section{Summary and conclusion}
\noindent
      In recent years, an impressive progress in the field of superheavy element search has taken place and heavy elements with Z~=~102~-~116 and 118 have been discovered at GSI, Darmstadt; JINR-FLNR, Dubna; and RIKEN, Japan. Some of these discoveries await final confirmation through heavy element chemistry. While searches for the ultimate long-lived neutron-rich superheavy element in the so-called, $island~of~stability$ are on, it is essential to find a reliable tool to predict their $\alpha$-decay half-lives to guide the experiments. 

      We present theoretical estimates for the $\alpha$-decay half lives of 314 heavy and superheavy elements with Z~=~102~-~120 in the WKB frame work with DDM3Y interaction, using available experimental and theoretical $Q$-values.  This formalism has been found to be quite reliable when experimental $Q$-values are used~\cite{prc06}. The theoretical $Q$-values are taken from the mass formulae of Myers-Swiatecki  [MS]~\cite{ms} and  Muntian et al. [M]~\cite{Mu01,MU03,Mu03}. The Viola-Seaborg-Sobiczewski [VSS] estimates of $\alpha$-decay half lives using the same  $Q$-values are also presented for comparison.

	It is pertinent to note that the experimental $Q$-values are not always reproduced by the theoretical ones extracted from the existing mass formulae and, calculations of $\alpha$-decay half lives are extremely sensitive to the choice of $Q$-values.  Incidentally, the $Q$-value predictions from Muntian et al. for the SHE domain and the corresponding $\alpha$-decay half lives are in better agreement with the available experimental data.  More over, half lives computed in the WKB frame work with DDM3Y interaction are found to provide a better description of the experimental data compared to VSS predictions with the same $Q$-values. 

The present calculation does not indicate a pronounced island of increased stability around Z~=~120.  In fact the half lives show a decreasing trend with increasing Z from Z = 114 to Z=120.  

Detail comparison with the experimental data available  so far in the heavy and superheavy mass region suggests that further improvement of mass formulae in the superheavy region is essential for more precise predictions of unknown half-lives. The need for improvements of mass formula having its origin based upon the basic nucleon-nucleon effective interaction is stressed. Ideally, a density dependent effective interaction such as the DDM3Y interaction which provides unified descriptions of the nuclear scattering, radioactivity and nuclear matter may itself be used for further improvements of the modified ~\cite{Mu01,MU03,Mu03} macroscopic-microscopic model.


\begin{figure}[h]
\eject\centerline{\epsfig{file=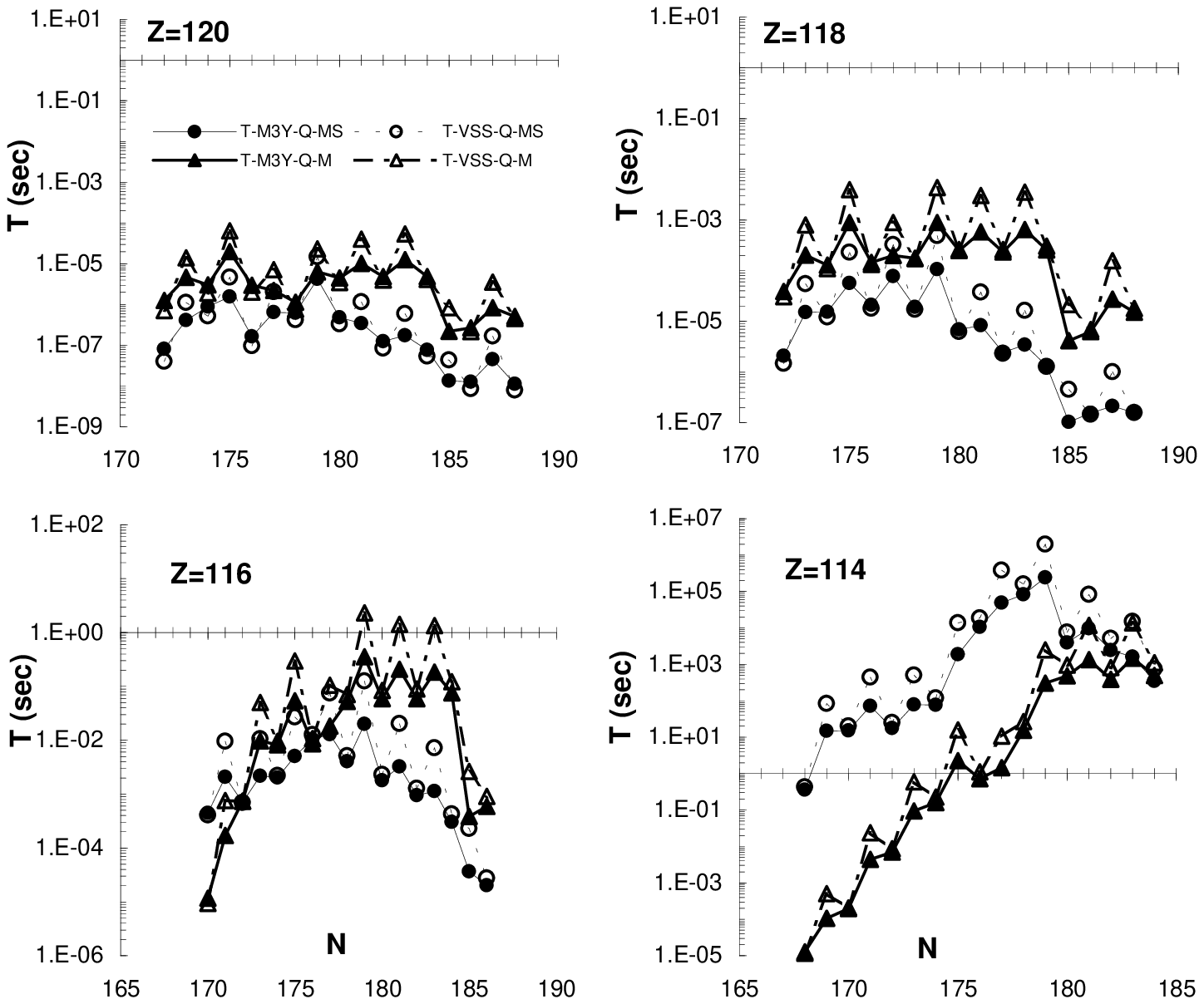,height=14cm,width=14cm}}
\caption
{Plots of alpha decay half-life [T(sec)] versus neutron number (N) for isotopes of Z=120, 118, 116, 114 calculated using theoretical Q-values from the mass formulae of MS [12] and M [16,17,18]. 
(a) DDM3Y with MS is represented by solid circle (T-M3Y-Q-MS), 
(b)VSS with MS is represented by hollow circle (T-VSS-Q-MS),
(c) DDM3Y with M is represented by solid triangle (T-M3Y-Q-M),
(d)VSS with M is represented by hollow triangle (T-VSS-Q-M).
The lines are guidelines to the eyes.}
\label{fig1}
\end{figure}

\begin{figure}[h]
\eject\centerline{\epsfig{file=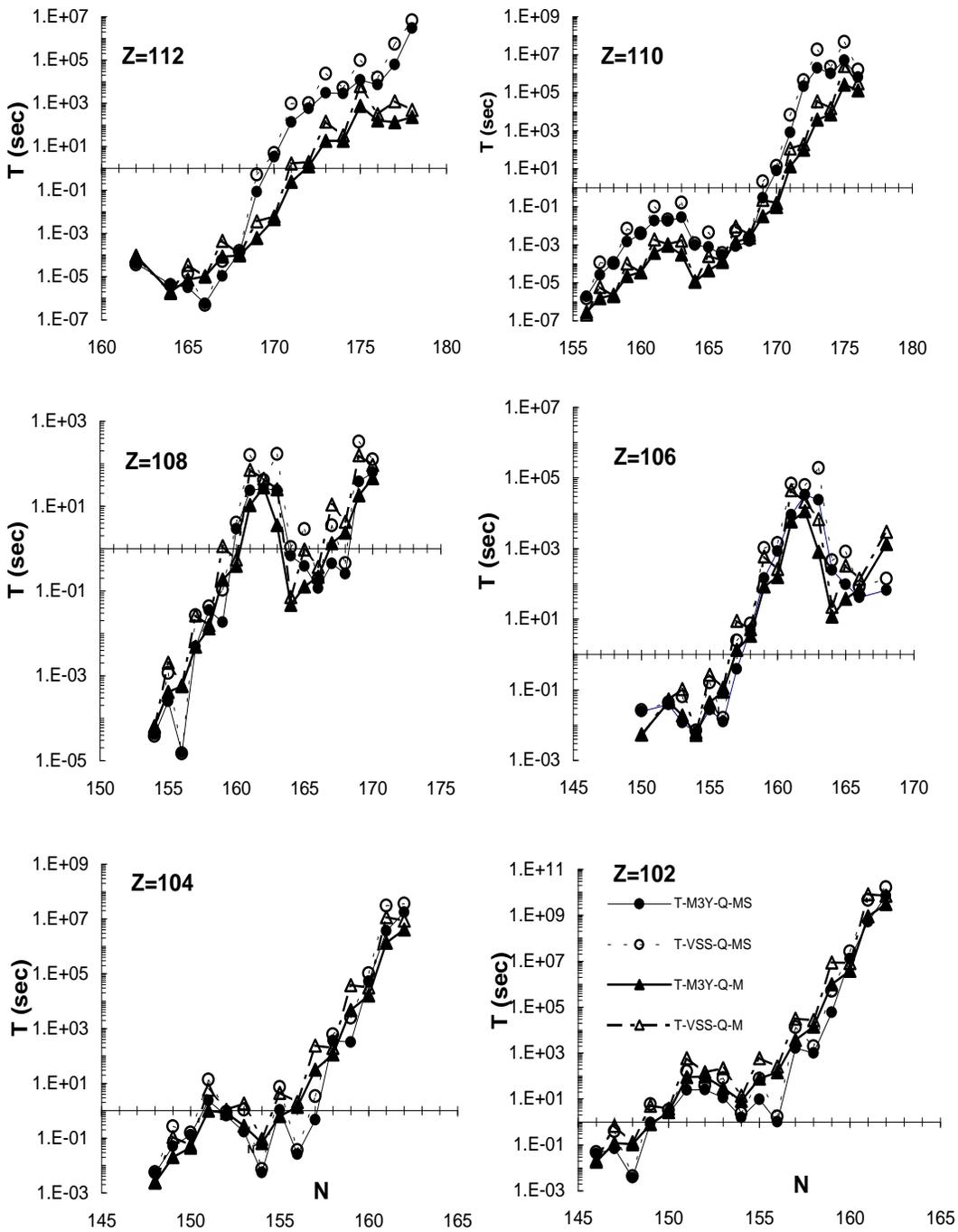,height=18cm,width=14cm}}
\caption
{Same as fig.1 for Z=112, 110, 108, 106, 104 and 102.}
\label{fig2}
\end{figure}

\begin{figure}[h]
\eject\centerline{\epsfig{file=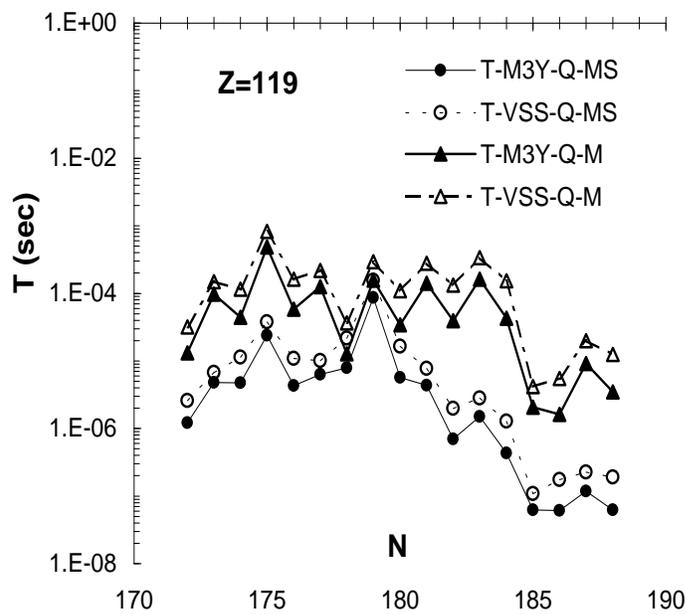,height=8cm,width=9cm}}
\caption
{Same as fig.1 for Z=119.}
\label{fig3}
\end{figure}

\begin{figure}[h]
\eject\centerline{\epsfig{file=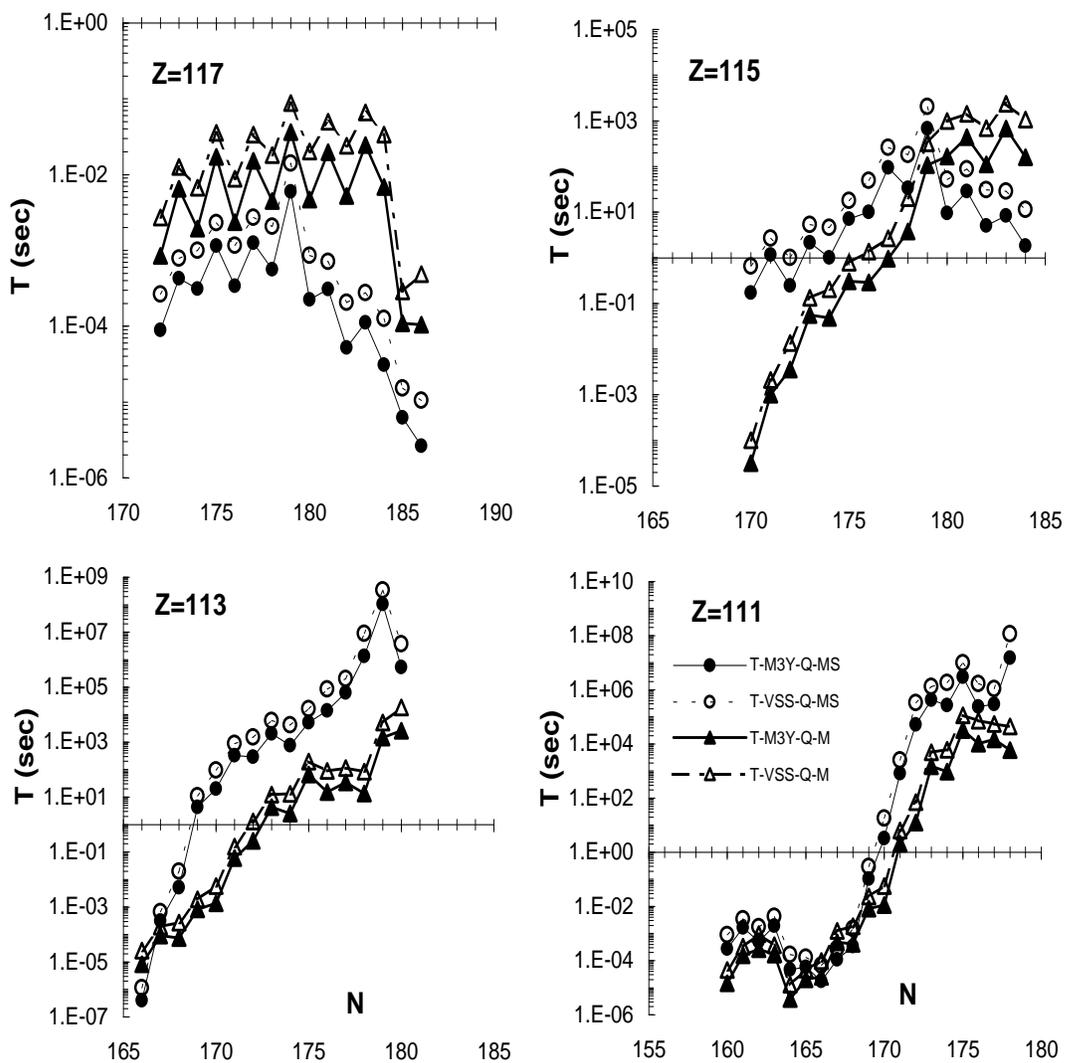,height=14cm,width=14cm}}
\caption
{Same as fig.1 for Z=117, 115, 113 and 111.}
\label{fig4}
\end{figure}

\begin{figure}[h]
\eject\centerline{\epsfig{file=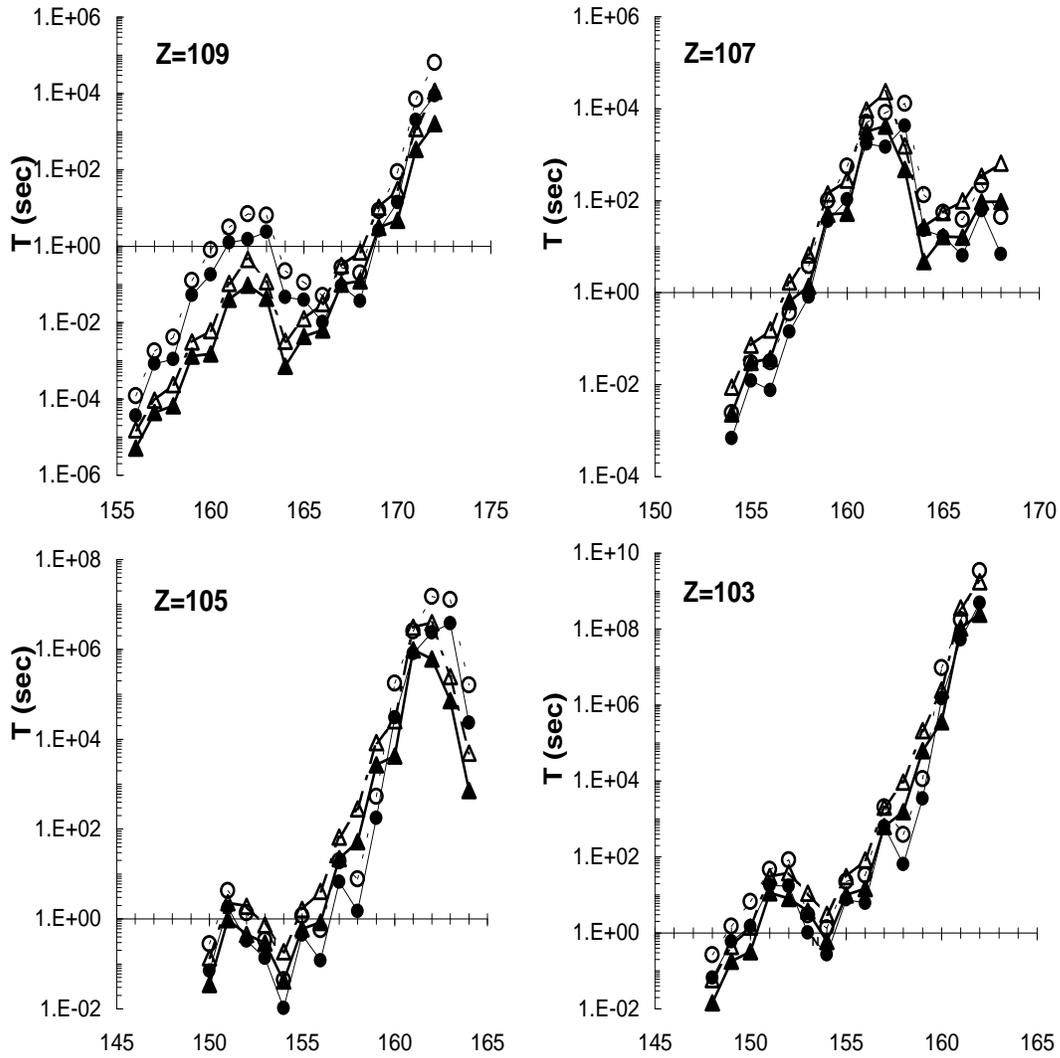,height=14cm,width=14cm}}
\caption
{Same as fig.1 for Z=109, 107, 105 and 103.}
\label{fig5}
\end{figure}


\begin{table}[tbh]
\caption{ Comparison of experimental [Exp] and calculated $\alpha$-decay half-lives using $Q$-values (MeV) from 
Exp~\cite{oga06,oga04,og04,mo04}, MS~\cite{ms} and M~\cite{Mu01,MU03,Mu03}.}
\begin{center}
\begin{tabular}{ccccccccccccc}
\hline \hline
 Parent &Exp&MS&M&Exp& This Work&This Work&This Work& Ref.  \\ 
\hline
$^A Z$&$Q_{ex}$&$Q^{MS}_{th}$&$Q^{M}_{th}$&$T_{1/2}$&$T_{1/2}[Q_{ex}]$& $T_{1/2}[Q^{MS}_{th}]$&$T_{1/2}[Q^{M}_{th}]$&Exp\\ \hline

$^{294}118$ &$11.81(6)$&12.51&12.11 & $0.89^{+1.07}_{-0.31}$ms&$0.66^{+0.23}_{-0.18}$ms&0.02 ms
&0.15 ms &\cite{oga06}     \\
$^{293}116$ &$10.67(6)$&11.15&11.09& $53^{+62}_{-19} $ms&$206^{+90}_{-61} $ms&12.8 ms&18.3 ms&\cite{og04}     \\  
$^{292}116$  &$10.80(7)$&11.03&11.06&$18^{+16}_{-6} $ms&$39^{+20}_{-13} $ms&10.4 ms&8.65 ms&\cite{og04} \\
$^{291}116$&$10.89(7)$&11.33&10.91&$18^{+22}_{-6} $ms& $60.4^{+30.2}_{-20.1} $ms&5.1 ms&53.9 ms&\cite{oga06} \\  
$^{290}116$&$11.00(8)$&11.34&11.08&$7.1^{+3.2}_{-1.7} $ms&$13.4^{+7.7}_{-5.2} $ms&2.0 ms&8.21 ms&\cite{oga06}  \\ 
$^{288}115$&$10.61(6)$&10.34&10.95&$87^{+105}_{-30} $ms&$410.5^{+179.4}_{-122.7} $ms&2161.6 ms&56.3 ms&\cite{oga04}  \\ 
$^{287}115$&$10.74(9)$&10.48&11.21&$32^{+155}_{-14} $ms&$51.7^{+35.8}_{-22.2} $ms&245.2 ms&3.55 ms&\cite{oga04}  \\ 
$^{289}114$& $9.96(6)$&9.08&10.04&$2.7^{+1.4}_{-0.7} $s&$3.8^{+1.8}_{-1.2} $s&1885 s&2.27 s&\cite{og04}     \\ 
$^{288}114$&$10.09(7)$&9.39&10.32&$0.8^{+0.32}_{-0.18} $s&$0.67^{+0.37}_{-0.27} $s&76.96 s
&0.16 s&\cite{og04}   \\ 
$^{287}114$&$10.16(6)$&9.53&10.56&$0.48^{+0.16}_{-0.09} $s&$1.13^{+0.52}_{-0.35} $s&77.74 s &0.09 s&\cite{oga06}  \\ 
$^{286}114$&$10.33(6)$&9.61&10.86&$0.13^{+0.04}_{-0.02} $s&$0.16^{+0.07}_{-0.05} $s&17.70 s &0.01 s&\cite{oga06} \\ 
$^{284}113$&$10.15(6)$&9.36&10.68&$0.48^{+0.58}_{-0.17} $s&$1.55^{+0.72}_{-0.48} $s&330.19 s &0.06 s&\cite{oga04} \\ 
$^{283}113$&$10.26(9)$&9.56&11.12&$100^{+490}_{-45} $ms&$201.6^{+164.9}_{-84.7} $ms&19845 ms&1.39 ms&\cite{oga04} \\ 
$^{278}113$&$11.90(4)$~$^{a)}$&12.77&---&344 $\mu$s~$^{b)}$&$101^{+27}_{-18}$$\mu$s&1.8 $\mu$s&---&\cite{mo04}  \\
$^{285}112$&$9.29(6)$&8.80&9.49&$34^{+17}_{-9} $s&$75^{+41}_{-26} $s&3046 s&18.6 s&\cite{og04}\\ 
$^{283}112$& $9.67(6)$&9.22&10.16&$3.8^{+1.2}_{-0.7} $s&$5.9^{+2.9}_{-2.0} $s &134.7 s&0.24 s&\cite{oga06}   \\ 
$^{280}111$&$9.87(6)$&10.34&10.77&$3.6^{+4.3}_{-1.3} $s&$1.9^{+0.9}_{-0.6} $s&0.10 s&0.01 s&\cite{oga04}  \\ 
$^{279}111$&$10.52(16)$&11.12&11.08&$170^{+810}_{-80} $ms&$9.6^{+14.8}_{-5.7} $ms&0.34 ms& 0.42 ms&\cite{oga04}  \\ 
$^{274}111$&$11.36(7)$~$^{a)}$&11.07&11.53& 9.26 ms~$^{b)}$&$0.39^{+0.18}_{-0.12} $ms&1.92 ms&0.17 ms&\cite{mo04}  \\ 
$^{279}110$&$9.84(6)$&9.89&10.24&$0.20^{+0.05}_{-0.04} $s&$0.40^{+0.18}_{-0.13} $s&0.29 s&0.03 s&\cite{oga06}\\ 
$^{276}109$&$9.85(6)$&10.11&10.09&$0.72^{+0.87}_{-0.25} $s&$0.45^{+0.23}_{-0.14} $s&0.09 s&0.10 s&\cite{oga04}  \\ 
$^{275}109$&$10.48(9)$&10.26&10.34&$9.7^{+46}_{-4.4} $ms&$2.75^{+1.85}_{-1.09} $ms&10.33 ms&6.36 ms&\cite{oga04}  \\ 
$^{270}109$&$10.23(7)$~$^{a)}$&9.73&10.27&7.16 ms~$^{b)}$&$52.05^{+27.02}_{-17.68} $ms&1235 ms&41.1 ms&\cite{mo04}   \\ 
$^{275}108$&$9.44(6)$&9.58&9.41&$0.19^{+0.22}_{-0.07} $s&$1.09^{+0.61}_{-0.35} $s &0.44 s&1.34 s&\cite{oga06} \\ 
$^{272}107$&$9.15(6)$&9.08&9.08&$9.8^{+11.7}_{-3.5} $s&$10.1^{+5.4}_{-3.4} $s&16.8 s&16.8 s&\cite{oga04}  \\ 
$^{266}107$&$9.26(4)$~$^{a)}$&9.00&8.95& 2.47 s~$^{b)}$&$5.73^{+1.82}_{-1.38} $s&36.01 s&50.8 s&\cite{mo04} \\ 
$^{271}106$&$8.67(8)$&8.59&8.71&$1.9^{+2.4}_{-0.6} $min&$0.86^{+0.71}_{-0.39} $min&1.59 min&0.64 min&\cite{oga06}   \\ 
\hline
\hline
\end{tabular} 
\end{center}
$a)$ calculated from $\alpha$-decay energies~\cite{mo04}; $b)$ experimental decay times~\cite{mo04}.\\ 
\end{table}

\end{document}